\documentclass[emulateapj,epsfig]{article} 
\usepackage{emulateapj}
\lefthead{Menci et al.} 
\righthead{} 

\begin{document}
\title{THE ABUNDANCE OF DISTANT AND EXTREMELY RED GALAXIES: 
THE ROLE OF AGN FEEDBACK IN HIERARCHICAL MODELS} 
\author{N. Menci, A. Fontana, E. Giallongo, A. Grazian, S. Salimbeni} 
\affil{INAF - Osservatorio Astronomico di Roma, via di Frascati 33, 00040 Monteporzio, Italy}
\smallskip 
\begin{abstract} 
The observed abundance of very red galaxies at high redshift has been recognized as a long standing 
problem for hierarchical models of galaxy formation. Here we 
investigate the effect of AGN feedback associated to the bright QSO phase onto the color 
distribution of galaxies from $z=0$ up to $z\approx 3$, and on the abundance of 
extremely red objects (EROs, with $R-K>5$) and distant red galaxies (DRGs, with $J-K>2.3$) 
up to $z\approx 4$; to this aim, 
we insert a blast-wave model of AGN feedback in our semi-analytic model 
of galaxy formation, which includes the growth of 
supermassive black holes and the AGN activity triggered by mergers and 
interactions of the host galaxies. In such a model, the AGN feedback is directly 
related to the impulsive, luminous quasar phase. 
We test our model by checking the consistency of its results against 
i) the QSO luminosity functions from $z=0$ to $z=4$; 
ii) the observed local relation between the black hole mass $m_{BH}$ 
and the mass (or velocity dispersion) of the host galaxy; iii) the color-dependent 
galaxy luminosity functions up to $z=3$.  
We then show how the efficiency of the AGN feedback associated to QSO phase 
increases with $z$. At low redshift it enhances the number of red bright
galaxies, so that the color distribution of $M_r<-22$ objects is 
entirely dominated by red ($u-r>1.5$) galaxies; at $0.5\leq z\leq 2$ it yields rest-frame 
$U-V$ color distribution in agreement with existing 
observations. In the range $z\approx 1.5-2.5$, we find that  $\approx 31 \%$ 
of galaxies contribute to the EROs population with $m_K<20$, in good agreement with the 
observed fraction $35\%$. 
In particular, at magnitudes $m_K=20$ (Vega system)
the model yields an EROs surface density of $6.3\,10^{3}$ deg$^{-2}$ matching existing data.
Extending our analysis to $z\approx 4$, the model matches the observed surface density $1.5\,10^3$ deg$^{-2}$ 
of DRGs at $m_K=20$; such a population 
is predicted to be dominated by galaxies with old stellar for $z\gtrsim 2.5$.
\end{abstract}


\section{INTRODUCTION}
The color distribution of galaxies constitutes a key observable to constrain
models of galaxy formation in a cosmological context. In fact, the most delicate 
sector of such models is that linking the age of stellar populations (closely 
related to the color) to the depth of the growing dark matter (DM) potential 
wells (hence the luminosity) of the galaxies hosting them. Indeed, while 
hierarchical models predict massive objects to be assembled at later cosmic times, 
they also predict their progenitor clumps to be formed in biased-high density 
regions of the primordial density field, where the enhanced density allowed 
early star formation. 

Thus, the abundance of red, massive galaxies resulting from  
such models results from the balance between the later epoch of assembly 
predicted for the most massive objects and the older stellar population 
predicted to be in place in the progenitors of such objects. 
Such a delicate balance makes the comparison with the observations 
concerning the abundance and the color distributoin 
of massive galaxies a sensible probe for such models.

Indeed, recent 
models, including the enhancement in the star formation of progenitors of 
massive galaxies triggered by interactions, are able to match many global 
properties of the evolving galaxy population, like the observed 
decline of the global stellar mass density from $z=0$ to $z=2$ (see, e.g., 
Fontana et al. 2004) and the evolution of the B and UV luminosity functions 
(Somerville , Primack \& Faber 2001; Menci et al. 2004a; Dahlen et al. 2005) from 
$z=0$ to $z=4$. 
The bimodal feature of the observed local color distribution (Strateva et al. 2001; Baldry et  al. 2004; 
Bell  et al.  2004; Giallongo et al. 2005) 
is also obtained in recent versions of hierarchical models 
(Kang et al. 2005; Menci et al. 2005; Bower et al. 2006; Croton et al. 2006), 
so that the existence of two well-defined populations appears to stem from the 
interplay between the biasing properties of the primordial density field -- 
originating the DM condensations constituting the progenitors of galaxies -- and the 
star formation and feedback processes driving the evolution of baryons in such progenitors.

Despite the above successes in matching the global behaviour of galaxy 
evolution, when one focuses on the observed proportion of red/blue galaxies, it 
is found that current hierarchical models underestimate the number of 
luminous/massive {\it red} galaxies. The models (Menci et al. 2005; Croton et 
al. 2006 with no AGN feedback) addressing the bimodal color distribution of 
local galaxies are characterized by an excess of massive ($M_r<-22$) galaxies in 
the blue ($u-r<1.5$) branch of the local color distribution compared to the red 
one; a similar deficit of luminous red galaxies is present in other semi-
analytic models of galaxy formation (Somerville 2004). At higher redshift the 
problem is even more severe: the abundance of extremely red objects (EROs, 
with optical infra-red colors redder than a passively evolving 
elliptical, $R-K>5$) at $z\approx 1-2.5$ is  underestimated by present 
hierarchical models (see McCarthy et al 2004; Cimatti et al. 2004; Daddi et al. 
2005; Somerville et al. 2004) by factors up to ten (depending on the exact 
color and magnitude cut). This suggests that an additional process must be at 
work in determining the observed properties of galaxies; the overall agreement 
of hierarchical model predictions with the global (not color-selected) evolutionary  
properties of galaxies discussed above suggests that such a process should 
not constitute the main driver, but rather a complementary physical 
mechanism affecting mainly the evolution of the bright galaxy population. 

In this respect, a possible solution has been proposed in terms of the feedback 
from Active Galactive Nuclei (AGNs) hosted at the center of galaxies (see, e.g., 
Ciotti \& Ostriker 1997; Silk \& Rees 1998, Haehnelt, Natarajan \& Rees 1998, 
Fabian 1999). Indeed, the role of the energy injection from the AGNs into the interstellar 
medium of the host galaxies is at present 
one of the most pressing issues in the study of galaxy evolution, both on the 
observational and on the theoretical side.

On the one hand, blueshifted absorption lines in the UV and X-ray spectra of 
active galaxies reveal the presence of massive outflows of ionized gas from 
their nuclei; they are characterized by high-velocities (up to 0.4 c) indicating 
mass flows of 1-10 $M_{\odot}$/yr (see Crensahw, Kramer, George 2003; Chartas 
2002, 2004; Pounds et al. 2003). Further evidences based on the radio and X-ray 
observations of galaxies (B\"oringer et al. 1993, Fabian et al. 2000; McNamara 
et al. 2000) indicate that bubbles of radio-emitting plasma are present in 
elliptical galaxies containing supermassive black <holes (SMBHs). On larger 
scales, fast (1000 km/s) massive (10-50 $M_{\odot}/yr$) flows of neutral gas are 
observed through 21-cm absorption of radio-loud AGNs (see Morganti, Tadhunter \& 
Oosterloo 2005), indicating 
that AGNs have a major effect on the circumnuclear gas in the central kiloparsec 
region of AGNs. Since the powers of outflows are similar to (or even excede) the 
observed bolometric luminosities and the cooling losses, these observations  
indicate that the feedback from AGN has to be considered among the processes 
which regulate the evolution of baryons in the 
galactic potential wells, and, of course, the growth of the SMBHs in the host 
galaxy. 

On the thoretical side, the impact of AGN feedback on galaxy formation has been 
investigated in a number of papers (Murray, Quataert \& Thompson, Monaco \& 
Fontanot 2005, Begelman \& Nath 2005; Scannapieco, Silk \& Bowens 2005); all 
the different assumed mechanisms for the energy injection can produce 
the expulsion of a significant fraction of the interstellar gas; the 
effectiveness of the AGN feedback in quenching the black hole growth and the 
subsequent star formation has been confirmed by recent aimed simulations of 
galaxy collisions triggering AGN activity (Di Matteo , Springel \& Hernquist 2005) 
and of AGN feeding activated by gas infall (Kawata \& Gibson 2005). 
However, inserting such a mechanism into a cosmological framework of galaxy formation 
constitues a challenging task. The model by Granato et al. (2004) successfully 
uses the shining of QSOs as a clock for the formation of elliptical galaxies, 
but does not include spiral galaxies, nor it provides predictions for the lower-
energy AGNs at $z<1$. In ''ab initio'' hierarchical models of galaxy formation 
in a cosmolgical context (Bower et al. 2005, 
Cattaneo et al. 2006; Croton et al. 2006; Kang, Jing, Silk 2006) SMBHs are 
assumed to grow during galaxy mergers both by merging with other BHs and through 
gas accretion,  the latter being described in terms of tunable scaling laws. The 
inclusion of AGN feeedback in these models allows to suppress the cooling in 
massive haloes (a long-standing problem of hierarchical models); however, 
such a feedback must be still at work at low redshift to continuously suppress 
star formation in massive haloes 
at $z\lesssim 1$. Since the QSO activity drops sharply for $z\lesssim 2.5$, 
these authors assume the energy feedback responsible for the suppression of the 
cooling to be effective only in halos that undergo quasi-static cooling, and is 
associated to a continual and quiescent accretion of hot gas onto the SMBHs. In 
these models the feedback mechanism is thus associated only to a smooth accretion process 
which is not the main driver of BH growth, so they do not focus on the evolution 
of the luminous properties of QSOs and of the bright AGN sources. 

Here we aim to investigate the effect on galaxy formation of the feedback 
directly associated with the observable QSOs and luminous AGNs. To this aim we 
include the AGN feedback model developed by Lapi, Cavaliere, Menci (2005) in our 
semi-analytic model of galaxy formation (Menci et al. 2004a; 2005), which 
includes the growth of SMBHs through gas accretion triggered by galaxy 
encounters. The AGN feedback we include in the present paper is substantially 
different from the implementations in Croton et al. (2006), Bower et al. (2005) 
and Cattaneo et al. (2006), since it is produced during the short AGN active 
phase, sweeping the cold gas content of the galaxies in a way similar to that 
resulting in the simulations by Di Matteo et al (2005); in addition, since it is 
associated with the observable active phase of AGNs, the feedback effect we 
investigate is mainly produced at high redshift. 
Before exploring the impact of such an impulsive 
form of AGN feedback on galaxy formation, we first test against observations the 
predicted evolution of the luminosity functions of QSOs from $z=4$ to the present, 
since the feedback we consider is directly related to their emission. 
After checking the consistency with the 
observed local $m_{BH}- \sigma$ relation between the BH mass $m_{BH}$ and the 
one-dimensional velocity dispersion $\sigma$ of the host galactic bulges, we compute the galaxy colour 
distribution at low ($z\lesssim 0.1$) and high (up to $z=3$) redshifts, 
discussing the effect of the AGN feedback on such distributions. 
Since the distinctive feature of our AGN feedback model (associating the feedback to the brigh
QSO phase) is constituted by an effect peaking at redshift $1.5<z<3.5$, we particulary 
focus on the properties of high-redshift galaxies, and we compare 
with the observed number density of EROs up to $z=2.5$, and of Distant red Galaxies (DRGs), selcted
by observed-frame $J-K>2.3$ (Franx et al. 2003; Van Dokkum et al. 2003), up to $z\approx 4$. 
In the final section we discuss our results.

\section{THE MODEL}
We adopt the semi-analytic model (SAM) of galaxy formation 
described in Menci et  al. (2005);  this connects the 
baryonic processes   (gas cooling,   star formation,   Supernovae feedback)  to 
the  merging  histories  of the  DM  haloes  (with mass $M$, virial radius $R$ 
and circular velocity $V$) and of the galactic sub-haloes (with mass $m$, radius $r$ 
and circular velocity $v$) following the canonical recipes 
adopted by SAMs of galaxy formation 
(Kauffman,  White \& Guiderdoni 1993; Cole et al. 2000; Somerville,  \& Primack  1999; 
Somerville, Primack \& Faber  2001;  Menci  et al. 2002; 2004a; Okamoto \& Nagashima 2003). 

The host DM haloes contain hot gas (with mass $m_h$) at the virial temperature $T$. 
The fraction of such a gas which is able to radiatively cool 
settles into a rotationally supported disk with gas mass $m_c$; the disk
radius $r_d$ and rotation velocity $r_d$ are computed after the model by Mo, Mao \& White (1998).
The merging histories of the host haloes, and the dynamical friction and 
binary aggregations acting on the included sub-halos, are computed adopting the 
canonical Monte Carlo technique as in Menci et al. (2005). We keep the 
implementation of gas cooling, star formation and feedback described in the 
above paper with the same choice of free parameters (the normalization of the 
star formation timescale and of the feedback efficiency). 

The integrated stellar emission from galaxies is computed convolving the 
star formation rate resulting from our model with the synthetic 
spectral energy distributions from Bruzual \& Charlot (1993), adopting a Salpeter 
IMF. The dust extinction affecting the intrinsic galactic luminosities is 
computed assuming the dust optical depth to be proportional to the metallicity 
of the cold phase (computed assuming a constant effective yield) and 
to the disk surface density (see Menci et al. 2002
for further details). When no otherwise specified, we assume a Galaxy extinction curve.
We have shown in 
our previous paper that the model is able to produce Tully-Fisher relation, 
cold gas and disk size distributions, and B-band galaxy luminosity functions 
(from $z=0$ to $z\approx 4$) in good agreement with observations. 
As for the growth of 
SMBHs with the associated the AGN emission, and for the corresponding AGN feedback,  
we adopt the models by Cavaliere \& Vittorini (2000) and by Lapi, Cavaliere \& 
Menci (2005), respectively. Here we recall their basic points. 

\subsection{The Growth of SMBHs}
The growth of SMBH is implemented as described in Menci et al. (2003); the 
accretion of cold gas is triggered by galaxy encounters (both fly-by and 
merging), which destabilize part of the avaiable cold gas. 
The rate of interactions is 
(Menci et al. 2003)
\begin{equation} 
\tau_r^{-1}=n_T\,\Sigma (r_t,v,V_{rel})\,V_{rel}. 
\end{equation}
Here $n_T$ is the number density of galaxies in the same 
halo and $V_{rel}$ is their relative velociyt. The encounters effective 
for angular momentum transfer require 1) the interaction time to be comparable with the 
internal oscillation time in the involved galaxies (resonance), ii) 
the orbital specific energy of the partners not to exceed the sum of their specific internal 
gravitational energies. The cross section $\Sigma$ for such 
encounters is given by Saslaw (1995) in terms of the distance 
$r_t\approx 2r$ for a galaxy with given $v$ (see Menci et al. 2003, 2004). 

The fraction of cold gas accreted by the BH in each interaction event is 
computed in terms the  variation $\Delta j$ of the specific 
angular momentum $j\approx Gm/v_d$ of the gas, to read (Menci et al. 2003)
\begin{equation}
f_{acc}\approx {1\over 8}\,
\Big|{\Delta j\over j}\Big|=
{1\over 8}\Big\langle {m'\over m}\,{r_d\over b}\,{v_d\over V_{rel}}\Big\rangle\, .
\end{equation}
Here $b$ is the impact parameter, evaluated as the average distance 
of the galaxies in the halo. Also, $m'$ is 
the mass of the  partner galaxy in the interaction,  and the average runs over 
the probability of finding such a galaxy in the same halo where 
the galaxy $m$ is located. 

The average cold gas accreted during an accretion episode is thus $\Delta 
m_{acc}=f_{acc}\,m_c$, and the duration of an accretion episode, i.e., the 
timescale for the QSO to shine, is assumed to be the crossing time for the 
destabilized cold gas component, $\tau=r_d/v_d$. 

The bolometric luminosity so produced by the QSO hosted in a given galaxy is 
then given by 
\begin{equation}
L(v,t)={\eta\,c^2\Delta m_{acc}\over \tau} ~. 
\end{equation}
We adopt an energy-conversion efficency $\eta= 0.1$ (see, e.g., Yu \& Tremaine 2002). 
The SMBH mass $m_{BH}$ grows through the 
accretion described above and by coalescence with other SMBHs during galaxy merging. 
As initial condition, we assume small seed BHs of mass $10^2\,M_{\odot}$ (Madau \& 
Rees 2001) to be initially present in all galaxy progenitors; 
our results are insensitive to the specific value as long as it is 
smaller than $10^5\,M_{\odot}$. 

In our Monte Carlo model, at each time step we assign to each galaxy 
the probability to interact after the rate given in eq. (1).
According to such a probability, we assign an active BH accretion phase (with 
duration $\tau$) to the considered galaxy, and compute the accreted cold gas and 
associated QSO emission through equations (2) and (3). 

\subsection{The Feedback from AGNs}

To explore the  dynamical effect of feedback occurring during the active AGN 
phase on the interstellar medium we adopt the model by Lapi, Cavaliere \& Menci 
(2005). They compute the effect of an energy injection $\Delta E$ by AGNs on the 
surrounding gas by solving, in the ''shell approximation'', the equations for a 
blast-wave propagating outwards in the interstellas gas, including the effects 
of gravity and of the gas density gradient. The perturbed gas is confined to a 
shell with outer (shock) radius $R_s(t)$ which sweeps the gas around the AGN; 
the effect is similar to that resulting in the simulations
by Di Matteo et al. (2005), although the latter have been performed only 
in selected cases of major mergers, while our treatment applies also to less
energetic events (the accretion rate, and hence the energy $\Delta E$ 
injected by the AGN, is determined by eq. 2). 

The mass $\Delta m$ expelled out of the virial radius by the blowout is computed 
as a function of the ratio $\Delta E/E$ where $E$ is the gas binding energy; 
The values of $\Delta m$ and $\beta$ for any given ratio $\Delta E/E$ are tabulated 
in Cavaliere, Lapi \& Menci (2002), the former being well approximated (to 
better than 10 \%) 
by $\Delta m/m\approx 0.5 \Delta E/E$ for $\Delta E/E<1.4$.

In our semi-analytic model, for any galaxy undergoing an active AGN phase (see 
sebsection 2.1), we compute $\Delta E=f\,L\,\tau$ assuming the AGN feedback 
efficiency $10^{-2}\lesssim f\lesssim  10^{-1}$ as a free parameter, the lower value being 
more appropriate to radio-quiet AGNs because of the flat spectrum and the low 
photon momenta; observations of wind speeds up to $v_W\approx 0.4 c$ suggest 
values around $v_w/2c\approx 10^{-1}$ associated with covering factors of order 
$10^{-1}$ (see Chartas et al. 2003; Pounds et al. 2003). We than compute the 
fraction $\Delta m$ of cold gas expelled by AGNs;  the galactic cold gas expelled 
by the AGN feedback enriches the hot gas phase which fills the dark matter potential 
wells of the structure (group or cluster) hosting the galaxy. 
Expanding out of the galaxy, the blastwave also expels a tiny fraction of the hot 
gas in the host structure and re-sets the hydrostatic 
equilibrium of such a hot gas to a new temperature $T+\Delta T$
larger than the initial temperature $T$ by a factor 
$1/\beta$ (with values in the range 1-1.1 in most cases),
also computed after Cavaliere, Lapi \& Menci (2002).

\section{RESULTS}
Here we present our results for a 
$\Lambda$-CDM cosmology with
$\Omega_0=0.3$, $\Omega_{\lambda}=0.7$, a baryon fraction
$\Omega_b=0.05$, and Hubble constant $h=0.7$ in units of 100 km
s$^{-1}$ Mpc$^{-1}$. The star formation and stellar feedback parameters 
are the same as in Menci et al. (2005), while the AGN feedback efficiency 
(see sect. 2.2) is set to $f=0.05$. 

\subsection{Testing the model: The AGN evolution and the Intergalactic Gas}
Since our feedback is tightly related to the active AGN phase, 
we require the model to match not only the observables concerning the galaxy population, 
but also those concerning the AGN population, in order to have a reliable 
modelization for the interplay between the cooling and star formation processes and  
the growth of SMBH with the ensuing AGN emission and feedback. 

\vspace{0.1cm} 
\begin{center} 
\scalebox{2.8}[2.8]{\rotatebox{0}{\includegraphics{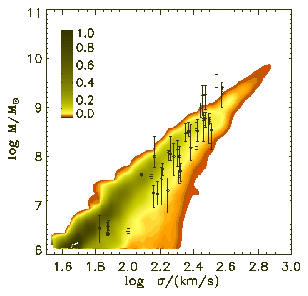}}} 
\end{center} 
{\footnotesize Fig.  1. - 
The relation between the black hole mass $M_{BH}$ and the 
one-dimensional velocity dispersion $\sigma$ of the bulges of the 
host model galaxies at $z=0$ is compared with data from 
Ferrarese \& Merritt (2000, filled squares) and Gebhardt et al. (2000, circles); 
the color code refers to logarithm of the abundance 
normalized to its maximum value. 
\vspace{0.1cm}
}

Thus, we first show in fig. 1 the $m_{BH}-\sigma$ relation, to check that, when 
integrated over time, the BH accretion that we implement in our model is consistent with 
the local observations for the whole range of galactic mass spanned by the 
model. Note that the spread in the $m_{BH}-\sigma$ increases for decreasing 
galactic mass, a feature shared by all hierarchical models and related to the 
increasing number of merging histories involved the formation of massive haloes, 
which results in a lower statistical deviation. Note also that our average 
relation remains close to a power-law, at variance, e.g., with the Bower et al. 
(2005) model, which predicts a steepening of the relation for large $\sigma$; in 
fact, such a steepening is due to the second ''mode'' of BH growth implemented in 
that model, associated with the smooth accretion of hot gas and which ensures 
that the luminosity of the BH is sufficient to quench cooling in the most 
massive haloes.

Since the time behavior of the energy injection from AGNs is crucial in 
determining the effects on the galaxy properties, and in our 
model such injection is  related to the active QSO phase, we test in fig. 2 the evolution of the 
QSO luminosity distribution by comparing it with the observations from $z=0$  
up to $z=4$. Note that our model is able to match the observed decline 
in the population of bright sources from $z=2$ to the present; this is due to 1) 
the decline of merging events refueling the cold gas reservoir of the galaxies; 
2) the exhaustion of such a reservoir, due to its earlier conversion in stars; 3) 
the decline in the rate of interactions (triggering the active accretion 
phase); 4) the decrease of the effectiveness of interactions in destabilizing the cold gas 
(see eq. 1 and the discussion in Menci et al. 2003).  
\vspace{0.1cm} 
\begin{center} 
\scalebox{0.37}[0.36]{\rotatebox{0}{\includegraphics{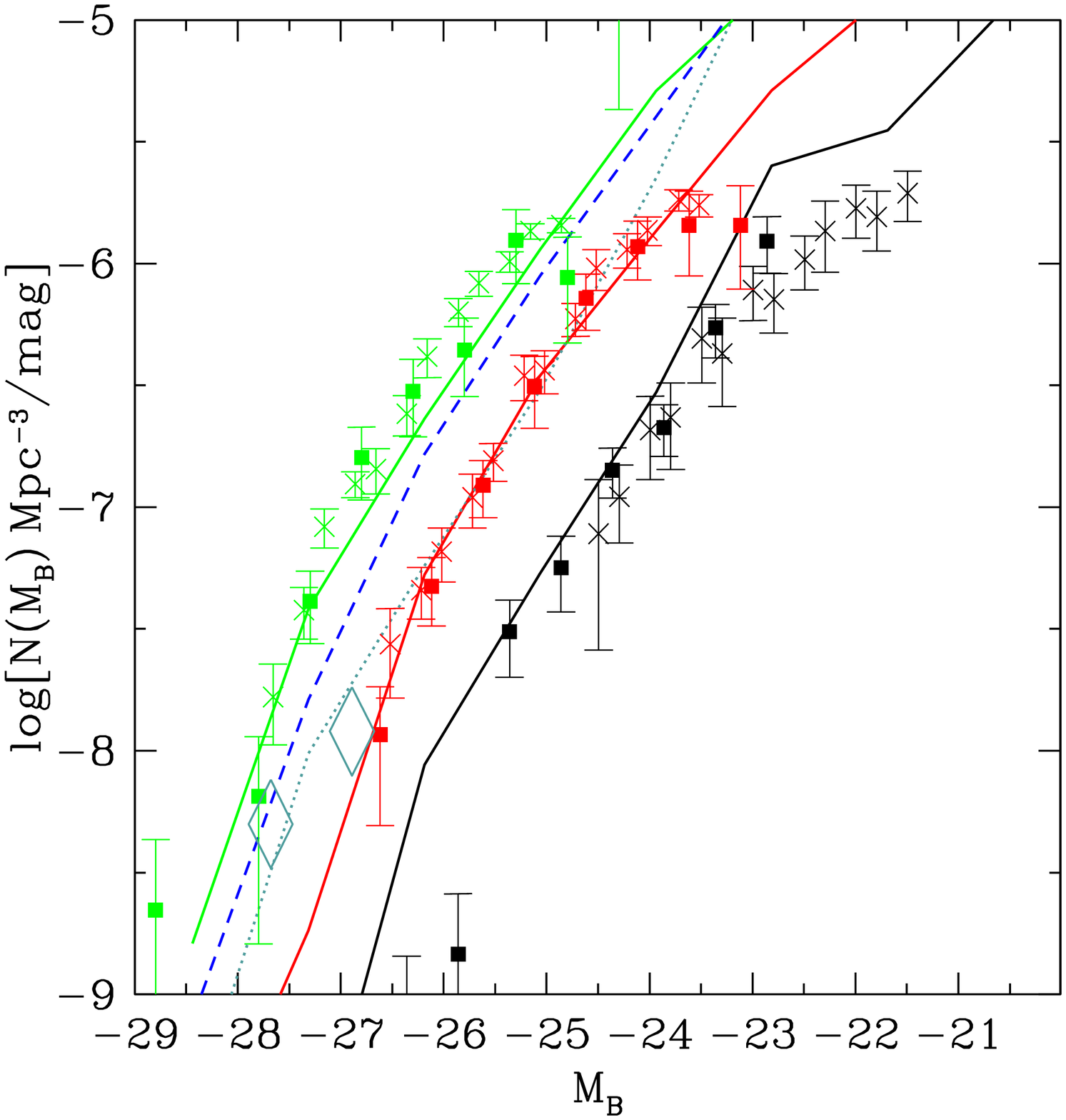}}} 
\end{center} 
{\footnotesize 
\vspace{-0.1cm} 
Fig.  2. - 
The QSO B-band luminosity functions 
from our model (solid lines) are shown for $z=0.55$ (lower curve), 
$z=1.2$ (middle curve) and $z=2.2$ (uppermost curve), and are compared with the 
data. These are taken from Hartwick \& Shade (1990, solid squares) and 
Boyle et al. (2000, crosses), and rescaled to our cosmology with $\Omega_0=0.3$, 
$\Omega_{\lambda}=0.7$, $h=0.7$. We also show the model results for 
$z=3$ (dashed line) and $z=4$ (dotted line), compared to the Sloan data at $z04.2$ 
from Fan et al. (2001, diamonds). 
The blue luminosity $L_B$ has been obtained by applying a 
bolometric correction of 13 (Elvis et al. 1994) to the bolometric luminosity in eq. (3). 
}

As a final test, we also probe the model predictions for the hot gas phase which is 
enriched by the gas expelled from the galaxies by the AGN feedback (see sect. 2.2). 
In particular we show in fig. 3 how the baryon fraction and the mass temperature 
relation resulting from the model compares with present observations. 
This shows that: i) that the 
amount of hot gas (enriched by the blowout of cold galactic gas due to 
the AGN feedback) is not enhanced over the baryon fraction in groups and clusters;  
ii) the temperature of the hot gas is not enhanced by the 
contribution from AGNs up to values exeeding the observations for a given 
cluster mass. Indeed, the value of $\beta$ (sect. 2.2) stays close to unity 
for typical values of the energy $\Delta E$ injected by AGNs. 
\vspace{0.cm}
\begin{center} 
\scalebox{3.2}[3.2]{\rotatebox{0}{\includegraphics{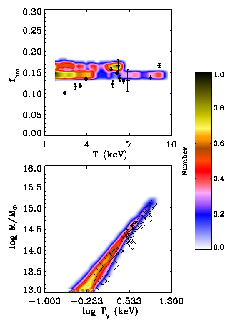}}} 
\end{center} 
\vspace{-0.2cm} 
{\footnotesize Fig.  3. - 
Upper panel: The relation between the baryon fraction and the X-ray temperature $T$ of the hot gas 
in groups and clusters of galaxies resulting from the model is compared with observational data points from 
Lin, Mohr, \& Stanford (2003); the color code refers to logarithm of the abundance of groups/clusters 
normalized to its maximum value. 
\newline
Bottom panel: The predicted relation between the mass $M$ of groups and clusters of galaxies and the 
X-ray temperature of the hot gas; the observational data points are taken from Figuenov, Reiprich \& B\"oringer 
}
\vspace{0.2cm}

The agreement with observations is not unexpected: indeed, while the AGNs 
are expected to produce significant effects on the cold gas contained in 
the shallow potential wells of galaxies (and hence to appreciably affect the 
star formation history of galaxies which entirely depends on the cold gas phase),  
the relative 
variation in the hot gas mass (which is an order of magnitude larger 
than the $m_c$) remains however small, and has little effect 
on the baryon fraction in groups and clusters of galaxies. In fact, 
for such structures the ratio $\Delta E/E$ (see sect. 2.2) is much less 
than unity due to their deep potential wells (large gas binding energy $E$), 
at least for our assumed value of the AGN feedback efficiency $f\approx 10^{-2}$, 
so that relative changes in the temperature ($\Delta T/T\lesssim 0.1$) and hot gas mass 
($\Delta m/m_h \lesssim 0.05$) remain much smaller than one.

\subsection{The Bimodal Color Distribution of Galaxies}

The effect of the energy injection from AGNs on the color distribution of local 
galaxies is shown in fig. 4 for three luminosity bins, and compared with the 
data from the Sloan survey (SDSS) given in Baldry  et al. (2004) for the $u-r$  colors; 
details on the SDSS $u$  and $r$ bands are given by the above authors. 
As expected, the AGN 
feedback affects only the color distribution of luminous galaxies; this is due 
to their larger cross section for interactions (triggering the AGN activity) and 
to the larger fraction of destabilized gas accreted by the SMBH (powering the 
AGN activity). Note also that both the {\it bimodal partition} and the {\it downsizing}
(i.e., the correlation bewetten the age of the stellar population, or the 
color, and the galaxy luminosity) are not determined 
by the AGN feedback, as we showed in our previuous paper and as confirmed by 
the recent results by Croton et al.(2006). Rather, the AGN feedback affects the partition of 
galaxies between the blue and the red population, enhancing the fraction of red galaxies. 
In particular, the distribution of  very luminous ($M_r<-22$) galaxies is dominated
by red objects when AGN feedback is included, in agreement with observations.   

\vspace{0.1cm} 
\begin{center} 
\scalebox{0.35}[0.35]{\rotatebox{0}{\includegraphics{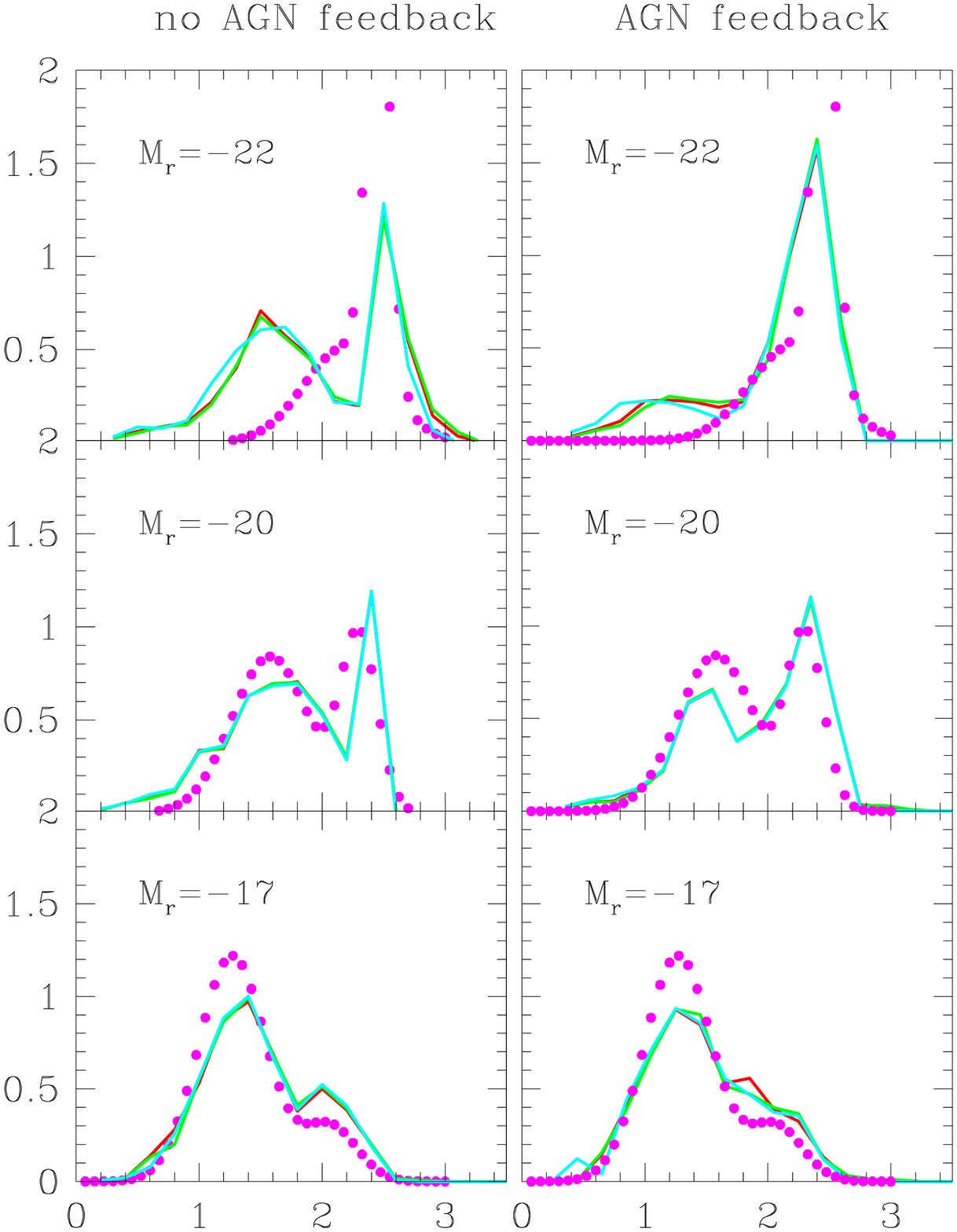}}} 
\end{center} 
{\footnotesize Fig.  4. - 
Predicted rest-frame $u-r$ 
color distributions (heavy lines) for different  dust extiction laws are 
compared with the  Gaussian fit to  the SDSS data  (from Baldry et al. 2004, 
dots) for different magnitude bins. The left column refers to the model without the
inclusion of AGN feedback, while right column shows the results when AGN feedback is turned 
on. The distributions are normalized to the 
total number of galaxies in the magnitude bin. 
\vspace{0.1cm}
}

At higher redshift, we expect the effect of the AGN feedback directly related 
to the bright QSO phase to be even more important, due to the enhanced AGN activity at 
such cosmic epochs. This is first shown in fig. 5, where we show the color-magnitude 
relation at various redshifts for the model with no AGN energy injection and the 
model where the AGN feedback is included. 
 
Note that the energy injection from AGN affects appreciably the number of red 
galaxies at redshifts $z\gtrsim 2$. In particular, at such redshifts, bimodality 
only appears in the model including the AGN feedback, while at lower redshifts
$z=1-2$ the bimodal partition of the color distribution is appreciably enhanced 
by the effect of AGNs.

\begin{center} 
\scalebox{3.5}[3.5]{\rotatebox{0}{\includegraphics{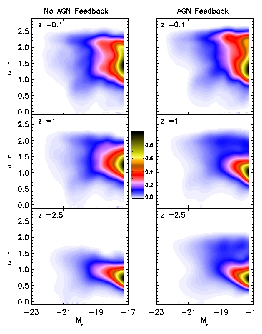}}} 
\end{center} 
\vspace{-0.2cm}
{\footnotesize Fig.  5. - 
The ($u-r$)-$M_r$ relations for the model with no AGN feedback 
(left column) and for the model including the AGN feedback (right column) 
at different redshifts. The color code refers to the number density of galaxies 
in a given color-magnitude bin, normalized to the maximum value attained at the
considered redshift.
}
\vspace{0.1cm}

To perform a more quantitative comparison between the predicted and the observed 
bimodal properties of the galaxy population at intermediate and high redshifts, 
we compare in figs. 6 and 7 with the observational results from the 
Great Observatories Origins Deep Survey (GOODS) database. 
We show in fig. 6 the B-band luminosity functions of the red and blue components at 
different redshifts and compare them with the GOODS data 
(Salimbeni et al. 2006). According to the selection criterium discussed in Giallongo et al. 
(2005), in this plot we define the blue and the red population as those separated 
by the rest-frame color $U-V=\alpha(19.9+M_B)+(U-V)_0$ (Vega system), where the parameters 
$ \alpha$ and $(U-V)_0$ are given in the caption for the three redshift bins. 
At high redshifts ($z\gtrsim 2$) the model with 
no AGN feedbak substantially underestimates the number of red objects, while 
the model including the AGN energy injection produces a reasonable agreement at all 
redshifts for both the blue and the red population, although the long-standing problem 
of the overprediction of faint objects is still present in the model. 
 
Note also how the galaxies effectively reddened by the inclusion of the AGN feedback 
constitute a minor fraction of the galaxies contributing to the B and UV band luminosity; 
although the abundance of bright galaxies in the UV and B bands
is indeed decreased by the inclusion of the AGN feedback 
(see top-left and bottom-left panels of fig. 6), the 
resulting luminosity functions are still consistent with the data. 

\vspace{-0.3cm}
\begin{center} 
\scalebox{0.43}[0.43]{\rotatebox{0}{\includegraphics{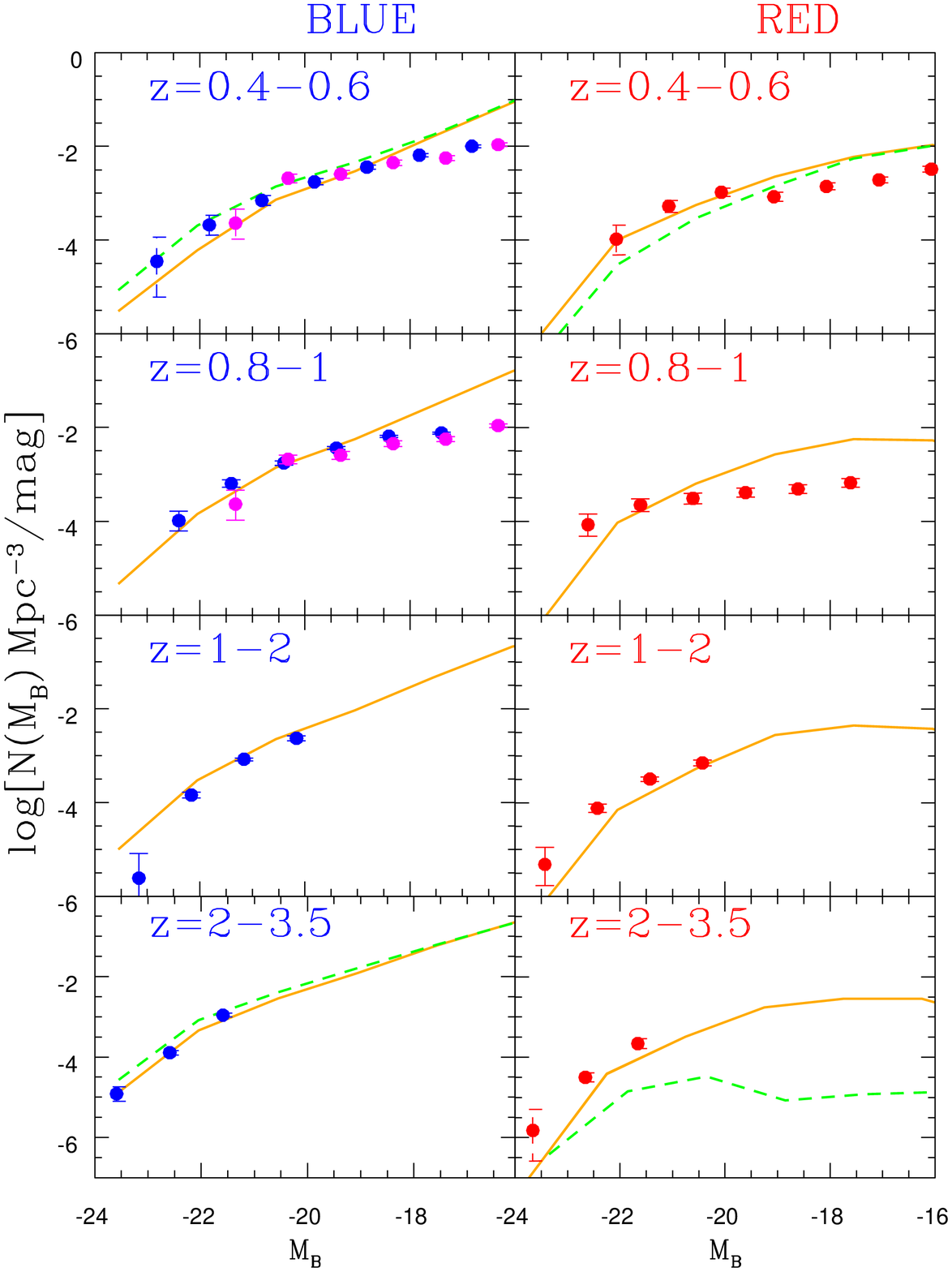}}} 
\end{center} 
\vspace{-0.5cm}
{\footnotesize Fig.  6. - 
Predicted B-band luminosity functions for the 
blue (left column) and the red (right column) population at different $z$
(solid lines) are compared with the GOODS data (Salimbeni et al. 2006).
The dashed lines show the model results when no AGN feedback is included. 
The parameters for the color separation between the two population 
(see text) are ; $\alpha=-0.084, -0.081, -0.083, -0.083$ 
for the $0.4-0.6$, $0.8-1$, $1-2$ and $2-3$ redshift bins, respectively; the corresponding values 
of $(U-V)_0$ are 0.66, 0.58, 0.53, 0.43. The Vega system is adopted for magnitudes.
}
\vspace{0.2cm}

A direct comparision of the predicted and the observed color distribution of 
galaxies at intermediate redshifts $z=1-2$ is perfomed in fig. 7 for three 
magnitude bins, to show how the model matches the observed early (at $z\approx 
2$) appearence of the same downsizing effect which marks the color distribution 
at low redshifts (see text below fig. 4), indicating that the model captures the 
observed {\it early} decline of the star formation rate in massive, bright 
galaxies. 

\begin{center} 
\scalebox{0.4}[0.4]{\rotatebox{0}{\includegraphics{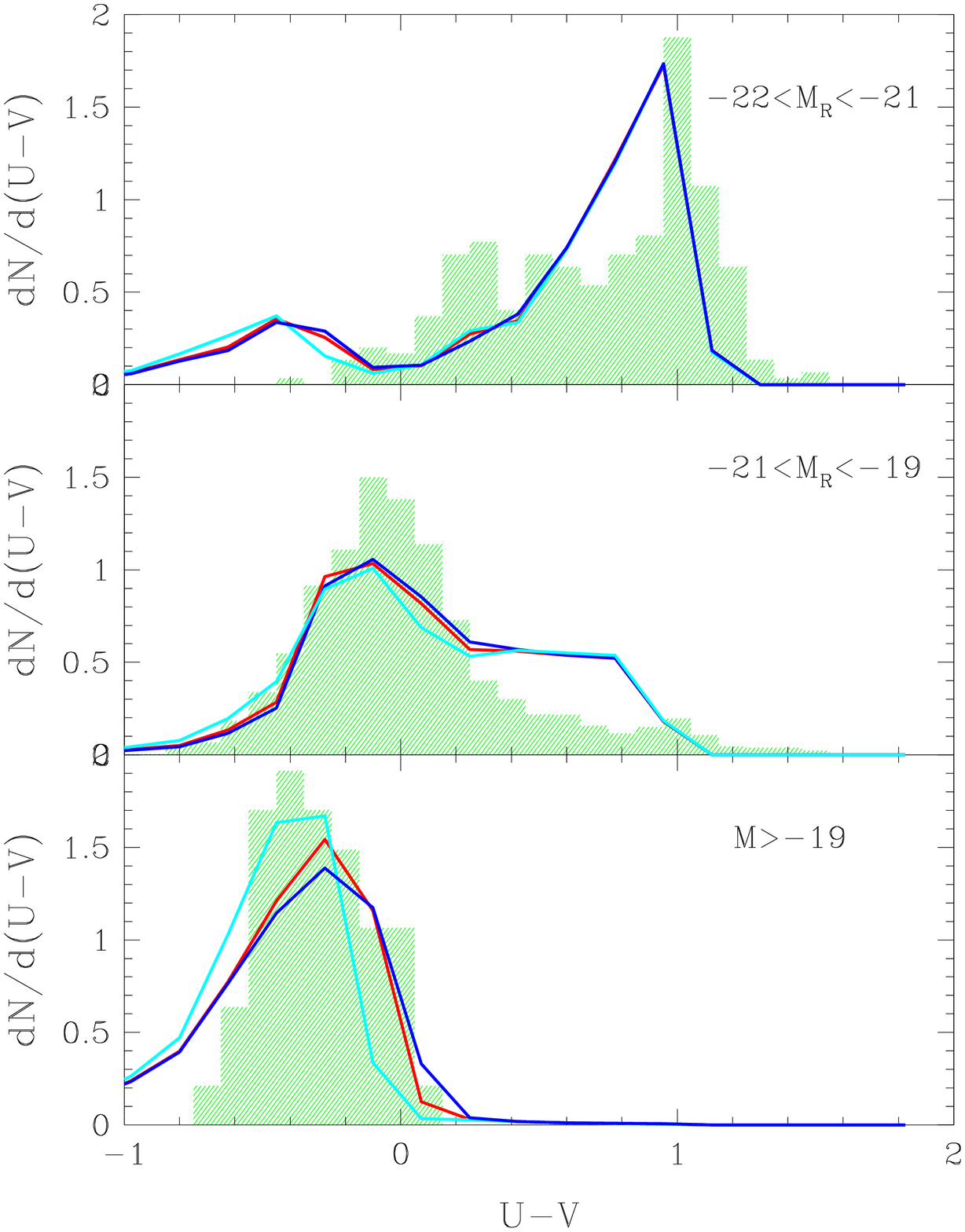}}} 
\end{center} 
\vspace{-0.2cm}
{\footnotesize Fig.  7. - 
Rest-fram U-V color distribution of galaxies in the redshift range 
$1<z<2$ for the R-band magnitude bins $M_r>-19$, $-21\leq M_r\leq -19$ and $-22\leq M_r<-21$.
The model predictions are the solid lines (for a Galaxy, SMC, or Calzetti extinction curves), 
while the histograms represent the GOODS data from Salimbeni et al. (2006).
}
\vspace{0.1cm}

\subsection{The Color Distribution and Abundance of EROs}
From the discussions above, and from inspection of fig. 5, we expect that the 
inclusion of the AGN feedback will solve the long-standing problem of semi-
analytic models related to their severe underestimate of EROs  (with observed 
$R-K>5$) at $z\approx 1.5-2.5$ (see McCarthy et al 2004; Cimatti et al. 2004; 
Daddi et al. 2005; Somerville et al. 2004)  that we have recalled in the 
Introduction. Such a point is addressed in fig. 8, where we plot the observed 
frame $R-K$ color distribution of $K<20$ (Vega system) galaxies in the redshift 
range $1.7<z<2.5$ and compare it with the observed distribution. 
Note how the inclusion 
AGN feedback strongly enhances the number of predicted EROs (see also fig. 5, 
bottom panel), to yield a fraction 0.31 of objects with $R-K>5$ close to the 
observed value 0.35. The 
normalization of both data and model predictions is provided by the redshift 
distribution of $K<20$ galaxies shown in the bottom panel. The same panel also
shows how the model matches the observed $z$-distribution of $m_K<20$ EROs,  
and (in the inset) the distribution of $I-K>4$ galaxies down to $m_K=21.5$ to probe the 
predictions of the model for the most luminous galaxies up to redshift $z=3$. 

Note that the matching between the model  {\it global} (not color selected) redshift 
distributions and the observations is not due to the effect of AGN feedback.
Indeed, this is due to the effect of starbursts triggered not only 
by margers (like, e.g., in the model by Somerville et al. 2004) but also 
by galaxy fly-by (see Menci et al. 2003); the latter produce starbursts with a lower efficiency 
$\approx 0.1-0.4$ (at $z\gtrsim 3$) for a single event, but dominate the encounter 
statistics due to the high rate of fly-by events at such redshifts; 
the point is discussed in detail in Menci et al. (2003).

\begin{center} 
\scalebox{0.4}[0.4]{\rotatebox{0}{\includegraphics{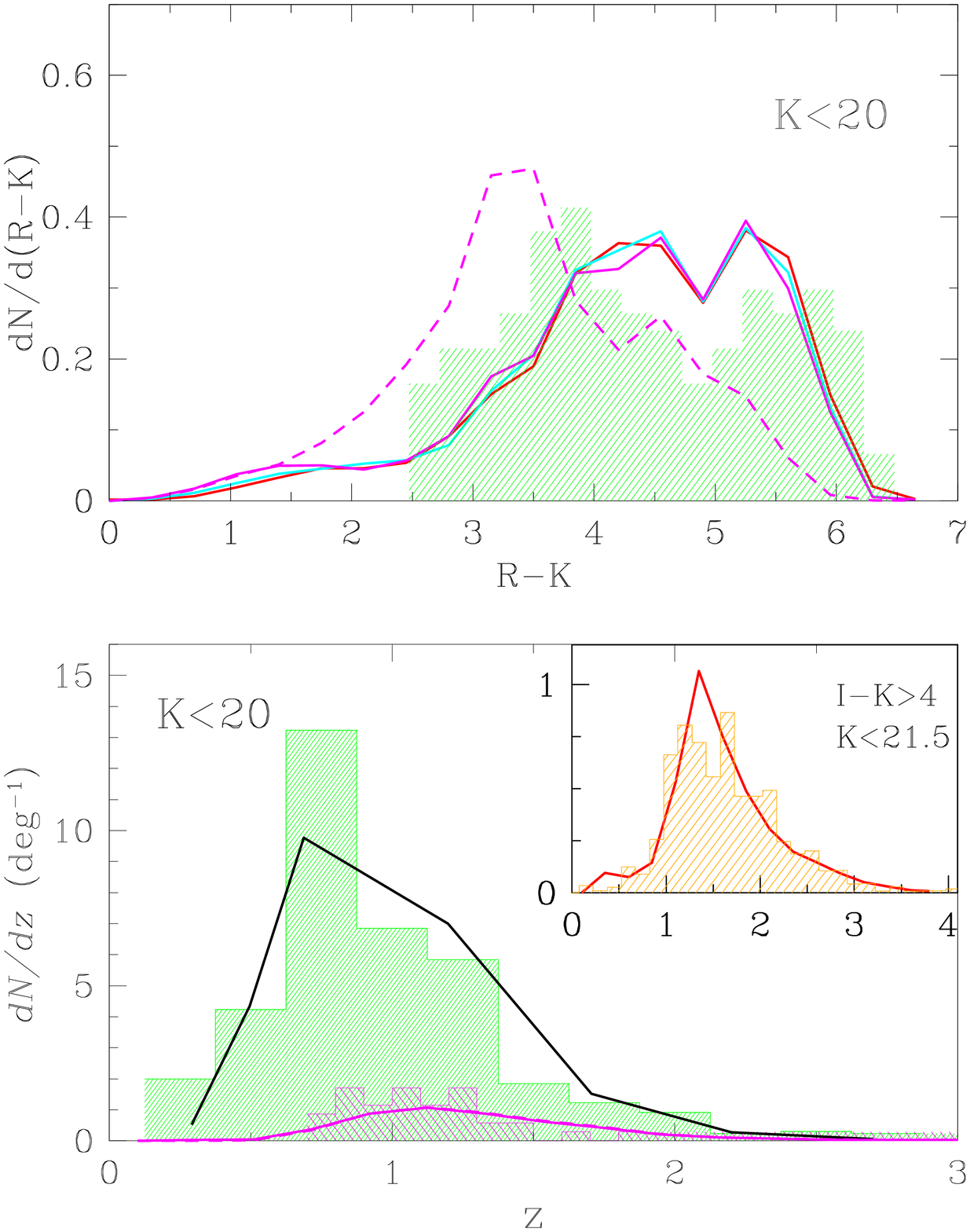}}} 
\end{center} 
\vspace{-0.3cm}
{\footnotesize Fig.  8. - 
Top panel: the observed-frame $R-K$ color distribution from the GOODS data (histogram, taken from 
Somerville et al. 2004) 
are compared with the results of the model with no AGN feedback (dashed line) and 
including the AGN feedback (solid line). \newline
Bottom panel: the redshift distribution of $K<20$ (Vega system) galaxies in the 
GOODS survey (histogram, taken from Somerville et al. 2004) are compared with the results of our model
(solid line). The lower histogram (taken from Cimatti et al 2002) and curve represent the observed and the 
model redshift distribution of the EROs population (normalized to the total number 
of objects for graphical reasons), respectively. 
\newline
The inset shows the redshift distribution of galaxies selected by $I-K>4$  
down to $m_K=21.5$; we compare with GOODS data from the catalog described 
in Grazian et al. (2006, histogram). 
}
\vspace{0.2cm}
The predictions in fig. 8 constitute a distinctive feature of 
our model, where the AGN feedback is associated with the bright AGN (QSO) phase 
(at variance with the other existing semi-analytic model, where it is associated 
to a smooth accretion phase, continuing down to low redshift), thus strongly 
acting on the properties of galaxies at $1.5\lesssim z<3.5$, when the AGN 
activity reaches its maximum. Thus we expect that in our model the density of 
EROs at magnitudes fainter than $m_K=20$ (contributed by high-redshift objects) 
to be still rising. Such a specific prediction of our model is shown in fig. 9
where we plot the differential magnitude counts of EROs resulting from our 
model. The model is able to match the observed value $\approx 6.3\,10^3$ deg$^{-2}$ 
for the surface density of EROs with magnitude $m_K\approx 20$, and predicts 
a continuous increase in the EROs surface density to reach values $\approx 
3\,10^4$ deg$^{-2}$ at $m_K=22$. 
 
The agreement of the model with observations 
holds not only for all EROs, but also in detail for density of dust-free passive EROs, 
which constitute about 80 \% 
of all EROs in our model. 
This is shown in the bottom panel of fig. 9, where we compare the model predictions 
for the cumulative number counts of EROs without dust extinction with the data concerning passive ù
dust-free EROs presented in by Miyazaki et al. (2003). 

\vspace{0.1cm}
\begin{center} 
\scalebox{0.35}[0.35]{\rotatebox{0}{\includegraphics{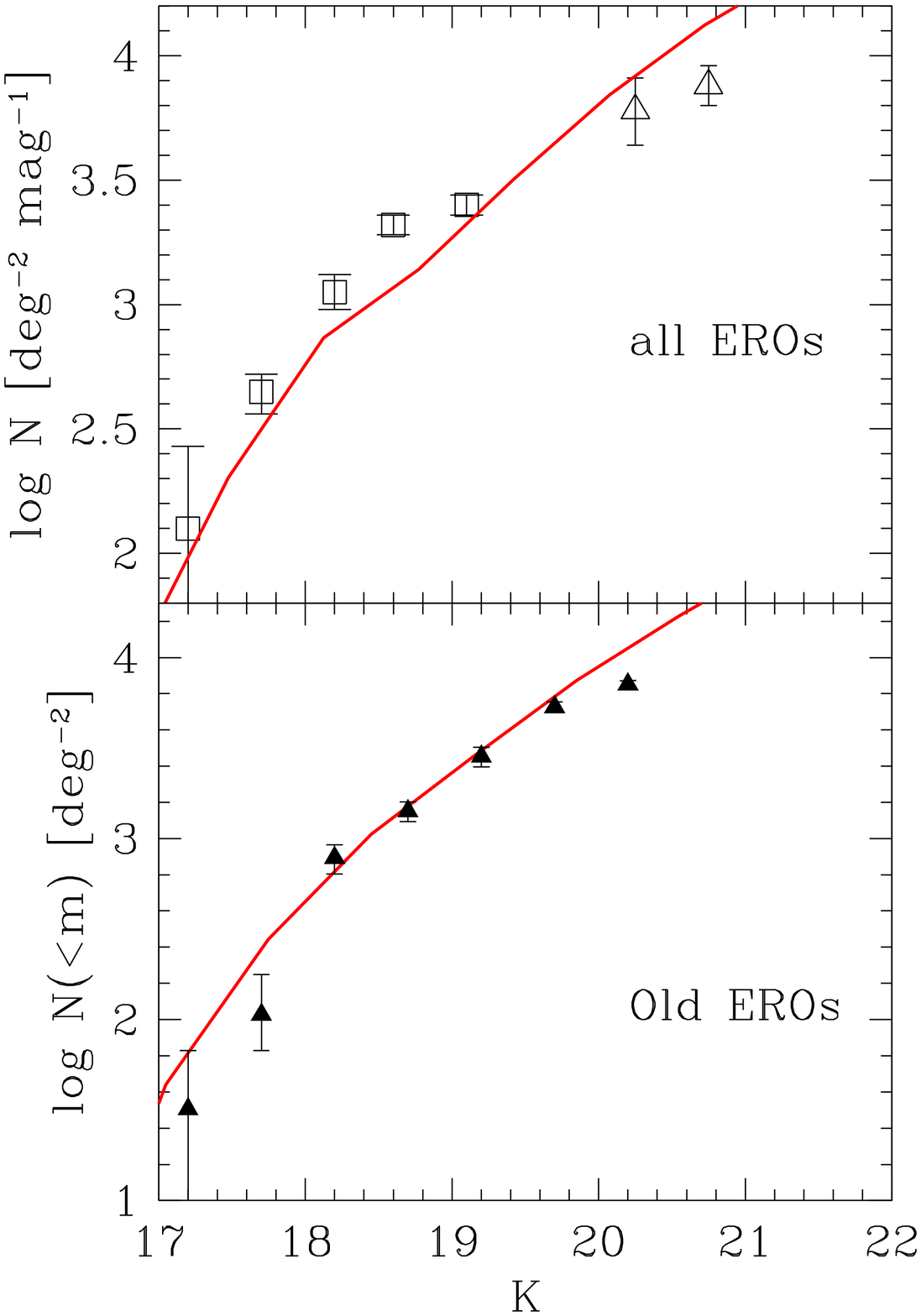}}} 
\end{center} 
\vspace{-0.2cm}
{\footnotesize Fig.  9. - 
Top panel: 
The surface density of EROs (per unit magnitude) as a function of their 
K-band magnitude resulting from the model is compared with data from Roche, Dunlop \& Almaini 
(2003, triangles) and Daddi et al. (2000, squares). \newline
Bottom panel: 
The {\it cumulative} surface density of EROs {\it without dust extinction} (solid line) is compared with data 
concerning the dust-free passive EROs taken from Miyazaki et al. (2003).
}
\vspace{0.1cm}

\subsection{The Abundance and Redshift Distribution of DRGs}
A more sensible probe for the model predictions at high redshifts is constituted by the observed 
surface density and redshift distribution of Distant Red Galaxies 
selected by the criterion $J-K>2.3$ (Franx et al. 2003; Van Dokkum et al. 2003), as 
shown in fig. 10. The predicted surface density of such objects (Top panel) matches the observed 
value  $1.5\,10^3$ deg$^{-2}$ at $m_K=20$, while at $m_K=22$ it reaches 
values $\approx 10^4$ deg$^{-2}$. Since the $J-K$ color cut allows for selection of red galaxies 
up to higher redshift compared to EROs, we can probe the model predictions concerning their 
redshift distribution up to $z\approx 4$ (bottom panel); the redshift distribution of 
the total DRG population is peaked at $z\approx 2.5$, in good agreement with the observed distributions, 
and shows a minor peak at $z\approx 1$. To investigate the nature of the DRG galaxies in our model, 
we also show the model predictions when no dust extinction is included (dashed line in the
bottom panel of fig. 10); in such a case, the low-redshift peak disappears, showing that the low-redshift part 
of the distribution is entirely contributed by heavily extincted galaxies peaking at 
$z\approx1$, while at $z\gtrsim 1.5$ the distribution is dominated by galaxies with old stellar population. 

\begin{center} 
\scalebox{0.4}[0.4]{\rotatebox{0}{\includegraphics{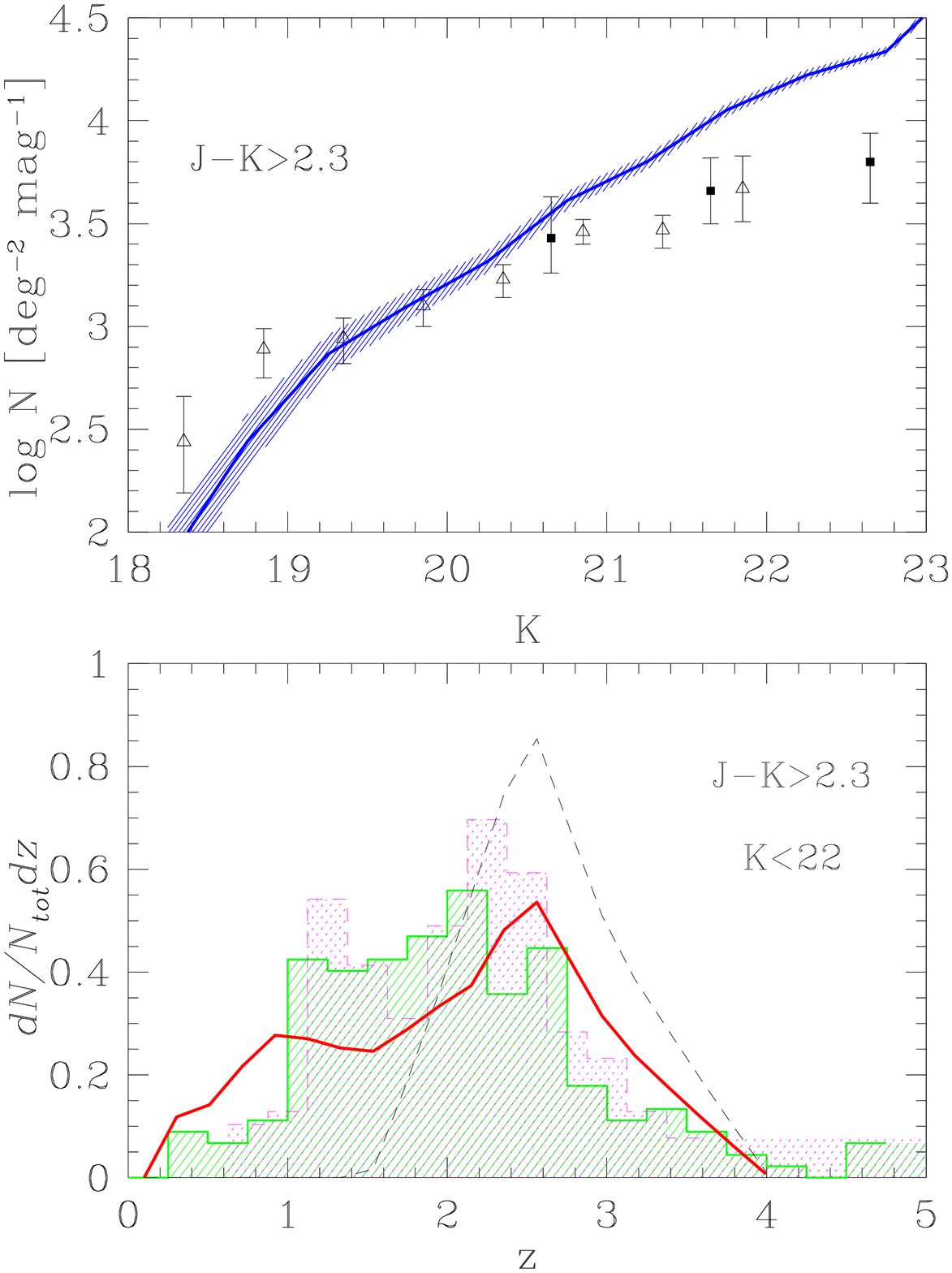}}} 
\end{center} 
\vspace{-0.2cm}
{\footnotesize Fig.  10. - 
Top panel: The prediced surface density (per unit magnitude) of DRGs is plotted as a function 
of their K-band magnitude (solid line). The dashed line shows the uncertainty due to the 
different adopted extinction curves. We compare with data from Grazian etal. (2006). 
Bottom panel: The predicted redshift distribution of DRGs with 
$m_K<20$ (solid line) is compared with the GOODS data from Grazian et al. (2006, solid histogram); 
we also show the HDF/Spitzer observations by Papovich et al. (2006, dashed histogram) 
for galaxies selected by stellar mass $m_*>10^{11}\,M_{\odot}$. 
The model prediction in absence of dust extinction are shown as a dashed line.
Both data and observations have been normalized to the total number of objects; their 
relative normalization is given by the magnitude counts in the top panel. 
}
\vspace{0.1cm}

\section{Summary and Discussion}

We have included the energy injection from AGN feedback in our semi-
analytic model of galaxy formation, which self-consistently includes the growth 
of SMBHs and the corresponding AGN emission. We have focussed on the effects of such 
a feedback on 
the evolution of the galaxy color distribution, and 
found that the inclusion of the energy feedback from AGNs enhances the number of 
red galaxies at low, intermediate and high redshifts, as to match existing observations 
up to $z\approx 4$. In particular, we find: 
\newline
$\bullet$ at low redshifts, the color distribution of bright ($M_r<-22$) galaxies 
is entirely dominated by red ($u-r>1.5$) objects (fig. 2). 
On the other hand, the color distribution of faint galaxies with ($M_r>-18$) remains 
peaked at blue ($u-r<1.5$) colors, and is not affected by the AGN feedback. 
\newline
$\bullet$ the effect of AGN feedback increases at higher redshifts, enhancing the 
fraction of red galaxies (figs. 4, 5, 6); a bimodal color distribution at $z\gtrsim 1.7$ 
is only obtained when the AGN feedback is considered (see the bottom panels in fig. 5). 
\newline
$\bullet$ at $1.5<z<2.5$ the model predicts a fraction 0.31 of EROs (with $R-K>5$) close to the observed 
value 0.35 (see fig. 8); the predicted surface density at $m_K=20$ is $6.3\,10^{3}$ deg$^{-2}$ while 
at $m_K=22$ it increases to $\approx 3\,10^{4}$ deg$^{-2}$ (fig. 9). \newline
$\bullet$ at higher redshift $2<z<4$ the surface density of predicted DRGs (with $J-K>2.3$) is 
$1.5\,10^3$ deg$^{-2}$ at $m_K=20$ in agreement with observations (fig. 10), while 
at $m_K=22$ the model yields values $\approx 10^4$ deg$^{-2}$, even slightly 
larger than current estimates $6\,10^3$ deg$^{-2}$ based on HST data. \newline
$\bullet$ the redshift distribution of $m_K<22$ DRGs is characterized by a major peak at $z\approx 2.5$ 
and a lower peak at $z\approx 1$ (see fig. 10); we find that the latter is contributed only 
by heavily absorbed galaxies, while at $z\gtrsim 1.5$ the distribution is dominated by 
galaxies with old stellar populations. Such a finding is in agreement with the recent results 
by Papovich et al. (2006), who measured the same partition in the redshift distribution of 
DRGs, although in their sample the peak of the lower-redshift, extincted galaxies 
is at a slightly larger $z\approx 1.5$. Our results are also consistent with the observed 
stronger clustering of high-redshift DRGs compared their lower-redshift counterpart (Grazian et al. 2006).  

We stress that in our model the downsizing (i.e., the older ages of stellar 
population observed in the most massive galaxies) and the appearence of a bimodal color 
distribution at $z\lesssim 1.5$ are not caused by the effect of AGNs. Indeed, 
hierarchical models predict massive objects to be assembled from progenitor 
clumps to be formed in biased-high density regions of the primordial density 
field, where the enhanced density allowed early star formation, so they 
naturally predict older stellar populations to be present in massive galaxies. 
In our model, the bimodality observed in the galaxy color distribution is the final results of 
the interplay between the above biasing properties of the primordial density field from 
which the progenitor of local galaxies have formed, and 
the dependence of feedback/star formation processes on the depth of the DM potential wells, as 
we showed in our previous paper (Menci et al 2005), and also obtained in independent 
works (Croton et al. 2006, case with no heating source). Rather, the AGN 
feedback strongly enhances the proportion of galaxies populating the red 
(or extremely- red) branch of the color distribution, and -- in our model --
such an enhancement is particulary effective at high-$z$ 
where the AGN activity is much larger than the present (see bottom panels in fig. 5). 

Note that the gas expelled from galaxies by the AGN feedback in our model enriches 
the hot gas content of the host halo (group or cluster); 
thus, the AGN feedback is partially counteracted by the cooling of 
such a hot gas which may re-convert part of it back to the 
cold gas at the center of galaxies. The final effect of the AGN 
feedback depends on the balance of the above two effects. 

At high redshifts, the AGN activity is so frequent, and the AGN 
luminosities so high, that AGN feedback rapidly counteracts such a 
cooling. The net result, is that galaxies retain - on average - a 
lower amount of cold gas (and hence show redder colors) compared to 
the case with no AGN feedback (see fig. 5, 
bottom panels). 

At lower redshift, the AGN activity decreases and the effectiveness of 
AGN feedback drops. Nevertheless, the effects of the AGN 
feedback (mainly generated at high redshifts) are not completely erased 
even at low redshift, since the reconversion of the hot gas in massive haloes
into the cold phase is suppressed by 
i) the lower gas densities in the larger host haloes 
(the radius of groups and clusters has grown); ii) the lower value of 
the cooling function (which drops for host halo virial temperatures 
$T>10^5$ K) haloes; iii) the reheating of cold galactic gas 
during major mergers as described in Menci et al. 
(2005). These processes (also present in our previous papers) 
allow for the persistence at low $z$ of effects (like 
the gas consumption in starbursts of by the AGN feedback) 
mainly active at high redshifts. 

Such a latter feature is peculiar of our model for AGN feedback, directly 
associated with the impulsive, luminous quasar phase (the main 
phase of BH growth) triggered by galaxy interactions. During such a phase, a 
small coupling to the gas is expected for the radiative output; estimates based 
on the observed wind speeds up to $v_W\approx 0.4 c$ suggest values 
$v_W/2c\approx 10^{-1}$ associated with covering factors of order $10^{-1}$, so 
that efficiencies close to our adopted value $f\approx 5\, 10^{-2}$ are 
expected. In this phase gas is swept out of the galactic potential wells in a 
way similar to that resulting from the simulation of Di Matteo et al. (2005). 
Such a mode of AGN feedback differs from that implemented in the other recent 
semi-analytic models, where the outflows are associated (with an efficiency 
close to 100 \%) to a lower-accretion phase of AGN, corresponding to accretion 
rates so low that the AGN is not optically luminous, and which is triggered by 
gas shock-heating (Cattaneo et al. 2006) or by smooth hot gas accretion within a 
static hot gas halo (Croton et al. 2006; Bower et al. 2005). The feedback 
associated to such a smooth mode of AGN accretion may start already at $z\approx 3$ 
but increases with time down to low redshifts. 
In this respect, the color distribution of galaxies at high-redshifts 
constitutes an important probe to discriminate between the two 
scenarios of AGN accretion and feedback pending a definite   
observational evidence for a statistically significant occurence 
of strong outflows in massive galaxies. 

\acknowledgments
We thank the referee for helpful comments.

\end{document}